# Molecular Dynamics Simulations of the $O_2^-$ Ion Mobility in Dense Ne Gas at Low Temperature: Influence of the Repulsive Part of the Ion-Neutral Interaction Potential

A. F. Borghesani
CNISM Unit, Department of Physics and Astronomy
University of Padua, Padua, Italy

and F. Aitken
University Grenoble Alpes, C.N.R.S., G2Elab
Grenoble, France

## ABSTRACT

New Molecular Dynamics simulations have been carried out in order to get an insight on the physical mechanisms that determine the drift mobility of negative Oxygen ions in very dense Neon gas in the supercritical phase close to the critical point. Two ion-neutral interaction potentials have been used that differ by their repulsive part. We have observed that the potential with a harder repulsive part gives much better agreement with the experimental data. The differences with the softer repulsive potential previously used are discussed. We propose that the behavior of the ion mobility as a function of the gas density is related to the number of neutral atoms loosely bound in the first solvation shell around the ion.

## 1 INTRODUCTION

THE formation of negative molecular ions as a consequence of electron attachment to electronegative molecules, such as $O_2$ or $SF_6$, is a process of paramount importance in many fields of chemistry and physics. Its influence spans industrial applications (e.g., plasma etching of semiconductors and insulation of high voltage equipments), environmental science via the physical chemistry of the ionosphere and the transport of ionized radicals in aerosols, as well as fundamental science in the field of ion transport in gaseous and liquid dielectrics. Actually, investigations on the drift mobility of negative ions in dense gases may help researchers shedding light on how ions fundamentally interact with the atoms of the host gas [1].

The measurements of the transport properties of ions in gases, which typically yield their drift mobility and/or diffusion coefficients (transverse and longitudinal) and allow the determination of the details of the ion-atom interaction potential, are usually carried out in conditions of extremely low gas density in order to get rid of the superposition of the potentials of neighboring atoms. Under these conditions, the study of the properties of stable anions is extremely difficult, as the excess kinetic energy of the attaching electrons leads to autodissociation of anions into charged- and neutral fragments [2,3].

On the other hand, the large collision rate of the newly formed anions with atoms in a high density, gaseous environment may lead to the stabilization of a significant fraction of ions that can then be drifted through the gas under the action of an externally applied electric field.

Unfortunately, only a handful of studies is concerned with the transport properties of anions in dense dielectric gases, which could establish the link between the limiting kinetic transport regime in dilute gas on one side and the hydrodynamic charge transport regime in liquids on the other one [4]. We believe that the reasons of this lack of detailed information may be ascribed to the difficulties of setting up an experiment that simultaneously satisfies the requirements of high pressure, low temperature, and high voltage at once and, more probably, to the lack of a full-fledged theory of ion transport at the intermediate densities of a dense gas that might have dampened the interest in such a topic.

At high densities the mean free path of the gas becomes comparable to the ion diameter and the binary collision picture, valid in the low density regime, breaks down. In the opposite situation of the hydrodynamic regime, in which the ion diameter is large compared to the atomic mean free path, the drag force on the ion and, consequently, its drift mobility is given by the well-known Stokes formula $\mu=e/6\pi\eta R$, where $\eta$ is the viscosity and $R$ the hydrodynamic radius.

In the intermediate density regime, several interpolating formulas have been proposed, the most famous of which is the Stokes-Cunningham formula, in which the deviations from the hydrodynamic Stokes law are proportional to the Knudsen number,

---

i.e., the ratio of the atomic mean free path to the ion radius. This approach, of some practical use, relies on the questionable assumption of what the hydrodynamic ion radius is. Typically, the Stokes-Cunningham formula is inverted to get an estimate of the ion radius, whereas the reverse would be a more satisfactory approach [5,6].

For positive He ions in superfluid He, however, a different, thermodynamic approach has been quite succesfully pursued. In this approach, the ion-fluid system is treated as a fluid mixture for which the van der Waals equation of state can be used. Then, the ion hydrodynamic radius is mainly determined by the van der Waals covolume of the ion [7].

Early studies on the $O_2^-$ ion mobility in dense gases were carried out in He [8,9]. More recently, we have investigated the behavior of $O_2^-$ in several noble gases in a broad range of densities and temperatures [10-12]. The results, however, did not provide a clear picture. We have shown that neither the kinetic theory nor the the Stokes law with a constant hydrodynamic radius can reproduce the experimental mobility data in the typical intermediate density range of a dense gas. Only in the proximity of the critical point in He and Ar the ion mobility can be explained by the hydrodynamic theory, provided that electrostriction is taken into account. The phenomenon of electrostriction leads to a large local increase of the density and viscosity around the ion that affects both its hydrodynamic radius and drag. Additionally, the long-range density fluctuations in the proximity of the critical point strongly affect the ion transport behavior by leading to the buildup of a correlated structure of gas atoms surrounding the ion, known as *solventberg*.

In particular, the measurements of the $O_2^-$ density-normalized mobility $\mu N$ as a function of the gas density $N$ in gaseous Ne for $T$=45 K are emblematic of the impasse in rationalizing the results. At relatively low $N$, $\mu N$ is roughly constant with a shallow minimum close to the critical density $N_c$=14.4 nm$^{-3}$ (the critical temperature is $T_c$=44.4 K). For larger $N$, $\mu N$ increases with increasing $N$ and saturates again to a constant value at very high density [10].

A general rationalization of these data is not presently available. At low $N$ the Kinetic Theory [13] is not able to reproduce the density dependence of the data even by using a scattering cross section derived by optimized ion-atom interaction potentials. Similarly, in the larger $N$ region the hydrodynamic Stokes formula [14] reproduces the data only if an adjustable, density-dependent hydrodynamic radius is introduced. Actually, the ion resides within an empty void surrounded by an local density enhancement that is the outcome of the competition between the short-range, repulsive exchange forces due to the Pauli exclusion principle and the long-range attractive polarization forces through which atoms and ions mutually interact. The density-dependent radius of the void is assumed to be the hydrodynamic radius to be used in the Stokes formula. This approach has given a quite good agreement with the experiment for the high-$N$ region but fails at lower $N$ because hydrodynamics is no longer valid there [15].

For these reasons we have carried out some Molecular Dynamics (MD) simulations for $O_2^-$ in Ar [16] and Ne [17] gas in order to investigate the microscopic mechanisms that lead to the observed mobility. In particular for Ne we have only obtained a qualitative agreement of the MD outcome with the experiment. As the repulsive part of the ion-atom interaction potential is not known, we have carried out further MD simulations with a different ion-atom potential. We report here the new results.

## 2 MD SIMULATION DETAILS

Details of the MD simulations have appeared elsewhere [16] and we briefly recall here their main features. We carried out equilibrium MD simulations that yield the ion diffusion coefficient $D$. The mobility $\mu$ is obtained by using the Nernst-Townsend-Einstein relation $\mu=k_BTD/e$. This procedure is justified because the experimental mobility has been measured in the limit of very weak field and the ions are in thermal equilibrium with the host gas. The simulated system is made of 1 ion and 100 Neon gas atoms. Once the interaction potential is given, the Newtonian equation of motions are integrated by implementing the Verlet algorithm [18]. The integration time step is $\tau$=2.42 fs. The initial configuration of the particles is chosen at random with the constraint that the velocities are distributed according to a Maxwell-Boltzmann equilibrium distribution. After $10^4$ equilibration steps, the system is allowed to evolve for $10^6$ more time steps for an overall simulation time of 2.4 ns. Every 500 time steps, i.e., every 1.2 ps, positions and velocities of all particles are recorded. Every 24 ps the velocities are simply rescaled so as to keep the temperature constant within 1 K [19]. This method has been tested by reproducing the well-known diffusion coefficient of pure Ar whose interaction potential is well known [20,21].

### 2.1 ATOM- AND ION-ATOM INTERACTION POTENTIALS

A crucial point for carrying out reliable MD simulations is the knowledge of the neutral-neutral and ion-neutral interaction potentials. The neutral-neutral potential is well known. A short-range repulsive Born-Mayer potential is added to the long-range attractive potential given by the damped dispersion series [21]

$$V_n(r) = A_{Ne}e^{-b_{Ne}r} - \sum_{n=3}^{5} f_{2n}(b_{Ne}r)\frac{C_{2n}}{r^{2n}} \qquad (1)$$

with $A_{Ne}$=199.5 hartree and $b_{Ne}$=2.458 bohr$^{-1}$ (1 hartree=27.2 eV, 1 bohr=529.2 pm). $C_6$=6.383, $C_8$=90.34, and $C_{10}$=1536 in atomic units (a.u.) are the dispersion coefficients [21,22]. The damping functions $f$ are introduced to prevent the divergence of the attractive part of the potential close to the ion and are given by

$$f_{2n}(x) = 1 - e^{-x}\sum_{k=0}^{2n}\frac{x^k}{k!}. \qquad (2)$$

Unfortunately, there are no literature references to the ion-neutral interaction potential. We have, thus, assumed a spherically symmetric potential given by

$$V_i(r) = A_i e^{-b_i r} - \sum_{n=2}^{3} f_{2n}(b_i r)\frac{D_{2n}}{r^{2n}}. \qquad (3)$$

The isotropy of the potential is justified by the small kinetic energy of the particles at low temperature that inhibits them to approach the ion too closely. The attractive polarization and dispersion forces are accounted for by the $D_{2n}$ coefficients. $D_4=\alpha/2 = 1.338$ a.u. where $\alpha$ is the atomic (dipole) polarizability of Ne. $D_6 = \beta/2+C'_6 = 25.87$ a.u. is related to the Neon quadrupole polarizability $\beta$ and to the $C'_6$ coefficient of the dispersion forces series expansion. The $C'_6$ coefficient is estimated according to the Slater-Kirkwood approximation [4].

The repulsive part of the ion-neutral potential is completely unknown. According to literature suggestions, it is assumed that it is proportional to the spatial distribution of the ionic charge [23,24]. We have computed it by using the Hartree-Fock-Roothan wavefunctions of the ion [25]. Some lines of constant charge density are depicted in Figure 1.

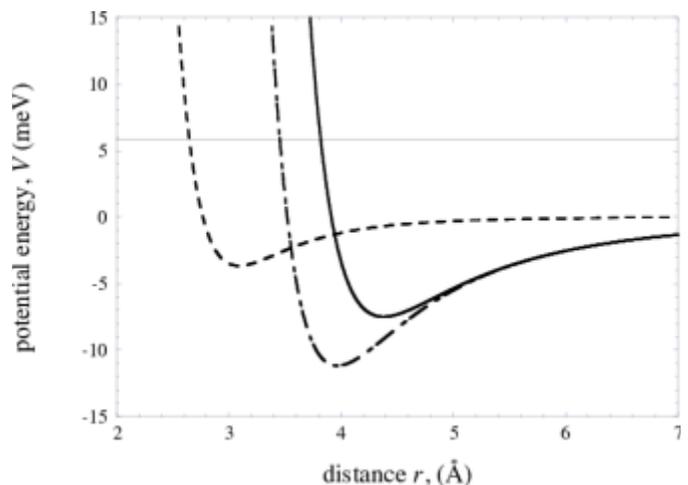

**Figure 2.** Ne-Ne (dashed line) and ion-neutral potentials. Solid line: potential hrp used in the present simulations with $A_i$=5000 a.u. Dash-dotted line: potential srp used in previous simulations with $A_i$=1630 a.u.

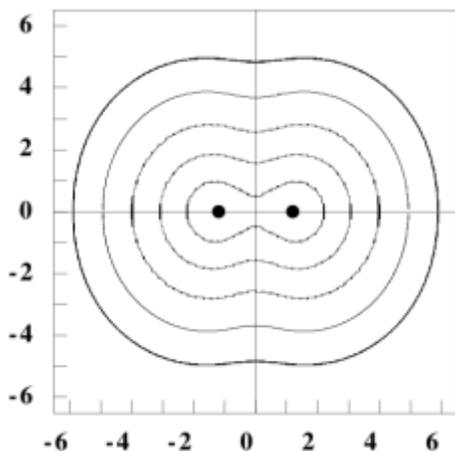

**Figure 1**. Some contour lines of constant electronic charge density $\rho$ of the $O_2^-$ ion. $\rho=1$, $10^{-1}$, $10^{-2}$, $10^{-3}$, and $10^{-4}$ a.u. (from the interior). 1 a.u. charge density = $1.08\ 10^{12}$ Cm$^{-3}$. The axes units are expressed in bohr.

By taking a spherical average of the charge distribution, we observe that the charge density is described by a decaying exponential of the form $e^{-b_i r}$ with $b_i = 2.154$ bohr$^{-1}$. Finally, the potential strength $A_i$ is completely unknown and it is left as an adjustable parameter.

A first attempt at determining the value of $A_i$ has been done by adopting a procedure by analogy. We first derived a rough estimate of the potential strength $A$ of the $O_2^-$-Ar system by inverting the dilute-gas mobility data for the systems $O_2^+$-He, $O_2^+$-Ar, and $O_2^-$-He [26]. Then, we inspected how the strength of the the $O^-$-rare gas atom potential changes as a function of the neutral species [27] and interpolated between the values of the previous estimates for the $O_2^-$. We thus obtained the value $A_i$ =1630 a.u. that yields a hard-sphere radius of 6.61 bohr = 0.35 nm and potential depth of $0.44\ 10^{-3}$ hartree = 12 meV. By using this value for the strength, the simulated $\mu N$ is roughly a factor 2 smaller than the experimental data [17].

This choice has evidently overestimated the well depth. Therefore, we adjusted the potential strength so as to agree with the experimental data at the highest investigated density, $N = 25$ nm$^{-3}$. By so doing we got the value $A_i = 5000$ a.u. that has been used throughout all new simulations. This new potential shows a hard sphere radius of 7.37 bohr = 0.39 nm and a depth of $0.273\ 10^{-3}$ hartree = 7.4 meV. The potentials used in all simulations are shown in Figure 2 together with the previous and the neutral-neutral ones. We term srp the previous softer repulsive potential and hrp the present harder repulsive potential.

## 3 MD RESULTS

In each run, 1 ion and 100 neutral atoms are simulated. Their motion is chaotic, typical of Brownian particles. The twodimensional projections of typical ion paths for densities $N = 5$ nm$^{-3}$ and 15 nm$^{-3}$ are shown in Figure 3. As expected, the range spanned by the ion at low $N$ is much larger than at higher $N$. The center-of-mass of the two simulations does not coincide because the initial configurations are chosen at random.

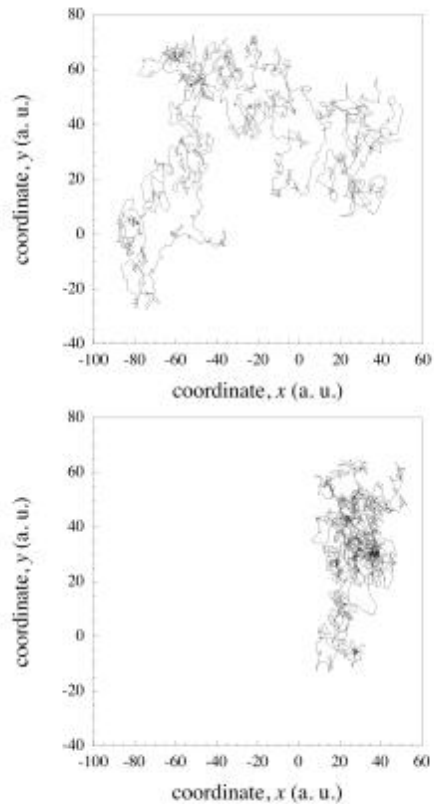

**Figure 3.** $x$-$y$ projection of the ion trajectories for $T = 45$ K and density 5 nm$^{-3}$ (top) and 15 nm$^{-3}$ (bottom). The axes units are a.u. (bohr).

The diffusion coefficient $D$ is obtained by a linear fit of the mean square displacement $\langle r^2 \rangle$ as function of time according to the equation

$$\langle r^2 \rangle = 6Dt. \quad (4)$$

As a consequence of the poor statistics of a single simulation, the time evolution of $\langle r^2 \rangle$ is quite erratic and the linear relationship predicted by (4) shows up only after averaging over a large number of simulations carried out in the same thermodynamic conditions. As an example, we show in Figure 4 the ionic mean square displacement for a few single runs together with its average over 100 similar simulations for $T = 45$ K and $N = 10$ nm$^{-3}$.

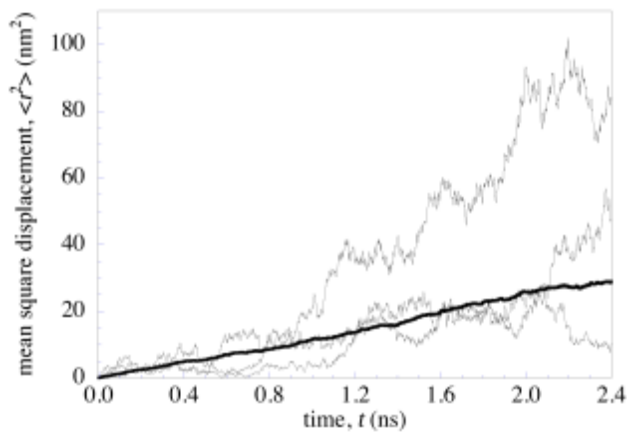

**Figure 4.** Thin lines: ion mean square displacement in three single runs. Thick line: its average over 100 runs. The simulation conditions are $T=45$ K and $N = 10$ nm$^{-3}$.

A linear fit to the average mean square displacement yields the ion diffusion coefficient and the density-normalized (or, reduced) mobility $\mu N$ is obtained as

$$\mu N = \frac{eD}{k_B T} N \quad (5)$$

In Figure 5 we show the reduced mobility $\mu N$ obtained from the present MD simulations and compare it with both the experimental data as well as with the outcome of the previous MD simulations carried out with the srp [17].

In comparison with the srp, the present one (hrp) has very much improved the agreement of the simulations with the experiment. For densities above the critical one, the simulations quite accurately reproduce the experimental data. At lower density the agreement with the experiment gets worse and the new simulations with the a harder repulsive potential give comparable results with the previous simulations with a softer repulsive potential.

The harder potential apparently gives origin to a minimum in the reduced mobility but its position and amplitude are very different from the experimental values. As we ascribed the existence of the mobility minimum to the proximity of the critical point [10], we note that this simple MD approach might not take into account the the long-range correlations that are present in the unperturbed fluid.

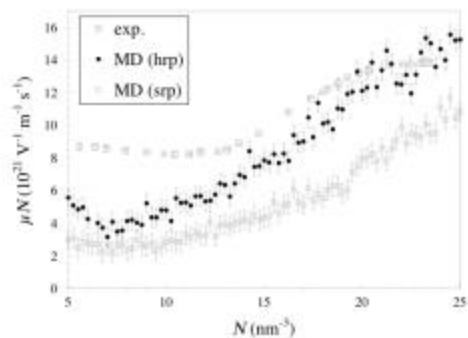

**Figure 5.** Density dependence of the density-normalized mobility $\mu N$ of $O_2^-$ in Neon gas for $T = 45$ K. Crossed squares: experiment. Closed points: MD simulations for the ion-neutral potential with repulsive strength $A_i = 5000$ a.u. (hrp). Open points: previous MD simulations with repulsive strength $A_i = 1630$ a.u. (srp).

### 3.1 DISCUSSION OF THE MD OUTCOME

The MD simulations do not allow the researchers to put into evidence the physical mechanisms responsible for the observed behavior of the mobility [19,28], although their agreement with the experimental data is quite good, especially at high density. In order to rationalize the MD results and to understand why the two potentials we used have given origin to such a different agreement with the experiment we have to investigate in great detail the physical quantities that the MD simulations compute.

The first of these quantities is the ion-atom pair correlation function $g(r)$ that is the probability to find a neutral atom at a given distance $r$ from the ion [18].

In Figure 6 we show the pair correlation functions obtained with the two different potentials for a density $N = 5$ nm$^{-3}$. Qualitatively they are very similar although their features depend on the respective potentials. In the $g(r)$ obtained with the hrp the position of the first peak is shifted to 9 bohr (0.48 nm) with the respect to the distance of 7.4 bohr (0.39 nm) of the srp. Moreover, the height of the first peak is now lower than previously obtained because the stronger repulsive part of the potential keeps the neutral atoms at larger distance. The $g(r)$ for the hrp case shows a minimum at a distance $r = 11.6$ bohr (0.61 nm) and a secondary maximum at $r = 13.1$ bohr (0.69 nm), whereas the same features occur at 10.2 bohr (0.54 nm) and 12. 2 bohr (0.65 nm) for the srp case.

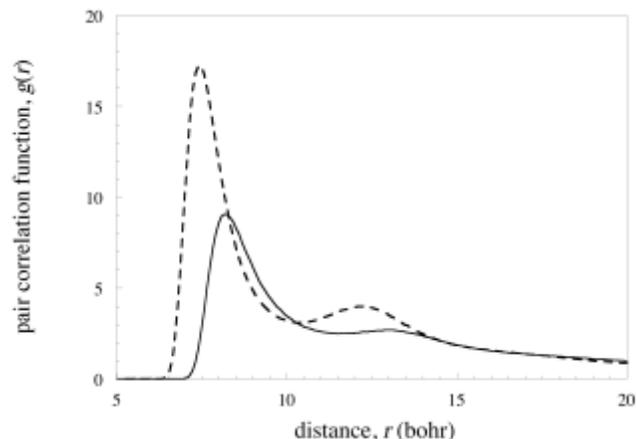

**Figure 6.** Ion-atom pair correlation function for $T = 45$ K and $N = 5$ nm$^{-3}$. Solid line: hrp. Dashed line: srp.

The presence of the minimum and of the secondary peak is a sign that a solvation shell is gradually growing around the ion. A dynamical exchange of atoms across the shell boundary is still possible as indicated by the nonzero value of the minimum. Moreover, the appearance of a secondary maximum suggests that also a second solvation shell is on the verge of growing up.

The first and second solvation shells turn out to be fully grown as the density is as high as $N = 19$ nm$^{-3}$, as can be observed by inspecting the behavior of the total correlation function $h(r) = g(r)-1$, shown in Figure 7, that measures the effect of one particle on the other due to the interactions [29]. At this density, $h(r)$ vanishes at the minima. For higher $N$, the values of the $h(r)$ minima even become negative.

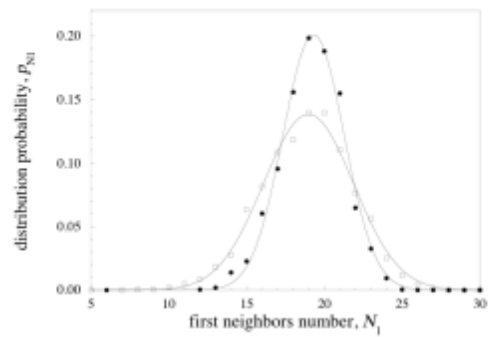

**Figure 8.** Probability distribution $p_{N1}$ atoms in the first solvation shell for $N = 15$ nm$^{-3}$. Closed points: hrp. Open points: rp..

density independent, but it is significantly smaller for the hrp than for the srp. We get $\tau_1 = (5.4\pm0.06)$ ps for the hrp and $\tau_1 = (7.9\pm0.1)$ ps for the srp. The ratio between the two values is 1.5 that approximately corresponds to the ratio of $\mu N$ obtained with the two different potentials.

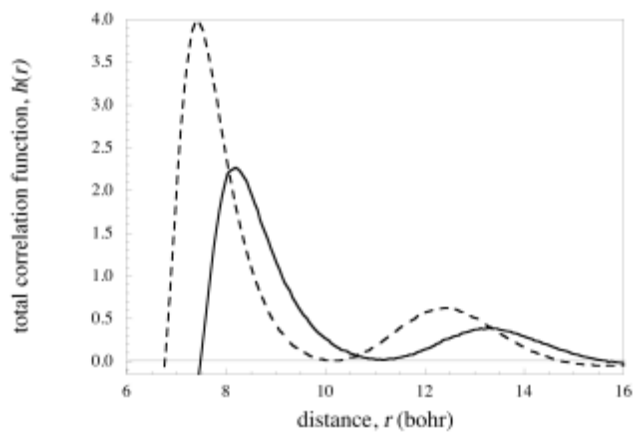

**Figure 7.** Total correlation function $h(r)$ for $N = 19$ nm$^{-3}$. Solid line: hrp Dashed line: srp.

Even at large densities, the different features of the correlation functions produced with the two potentials remain qualitatively the same. In particular, the harder potential keeps correlating less neutral particles than the softer one.

The motion of the ions should be mostly affected by the nearest neighboring atoms that are located within the first solvation shell. Their probability distribution is described by a Gaussian function and is shown in Figure 8 for $N = 15$ nm$^{-3}$. The average number of nearest neighbors $N_1$ is roughly the same for both potentials used but the variance is a bit larger for the hrp. The distribution of the nearest neighbor residence time within the range of the first solvation shell is exponential, as it is to be expected for a memoryless Markovian process such as the Brownian motion [30].

The average number of neutral atoms that are to be found within a distance from the ion corresponding to the radius of the first solvation shell is a weakly increasing function of the gas density, as shown in Figure 9. At lower $N$ the hrp correlates less atoms than the srp but the trend is reversed at higher $N$. Owing to the width of the nearest neighbor probability distribution, we can conclude that the ion is surrounded, on average, by the same numbers of neutral atoms, independently of the choice of the ion-atom potential.

What is really different for the two potentials, however, is the average residence time $\tau_1$ of neutral atoms within the first solvation shell. For both potentials it is roughly constant,

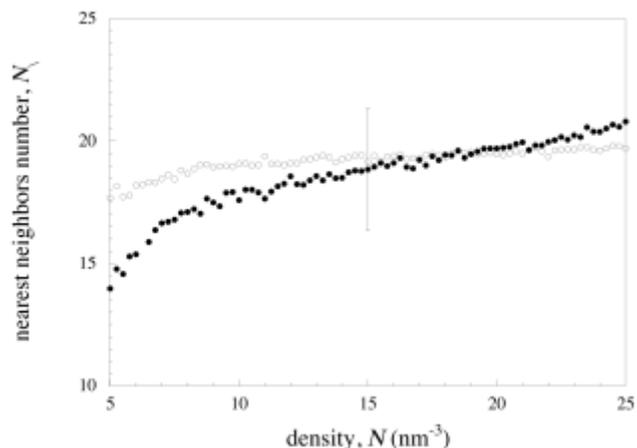

**Figure 9.** Density dependence of the average numbers of nearest neighbors $N_1$ for $T = 45$ K. Closed symbols: hrp. Opens symbols: srp. The error bar is a measure of the distribution width.

## 4 CONCLUSIONS

Equilibrium Molecular Dynamics is a well-established computational technique exploited in several branches of Physics, Chemistry, and Biology. In particular, it appears to be the only reliable tool to investigate the transport behavior of an ion in an intermediately dense environment, such as a dense gas, that is at the same time too compressed for Kinetic Theory to apply and too dilute for the validity of a full hydrodynamic treatment.

If a limited number of particles is to be simulated, MD is easily run even on personal computer and provides the researchers with a great number of physical interesting properties including transport coefficients and correlation functions. In the present case of a $O_2^-$ ion in dense, cold Neon gas, we have shown that the optimization of the ion-atom interaction potential leads to a great improvement of the MD simulation results as far as the ion diffusion and mobility are concerned.

The simulations demonstrate that the ion is gradually correlated with more and more atoms as soon as the density increases until, at high density, a fully developed first, and very probably also a second, solvation shell appear. The two ion-atom interaction potentials we have used give qualitatively similar results with the most important difference that the average residence time of neutral atoms in the immediate proximity of the ion is shorter for the more repulsive potential.

How all these observations could be used to rationalize the mobility results is still unclear. On one hand, the weak increase of the average nearest neighbors number with increasing density should lead to a slow decrease of the mobility $\mu$ in a so broad density range that the product of mobility times density $\mu N$ increases with increasing density. On the other hand, it seems that a reduction of the average time during which the ion can interact with the neutral atoms in its immediate proximity within the first shell leads to an increase of the mobility itself. As a result, the harder is the repulsive potential, the higher is the reduced mobility of the ion.

We believe that further refinements of the ion-neutral atom potential should improve the prediction of the transport properties of the ion. However, we also believe that there still is room for improving the neutral-neutral potential to better reproduce the thermodynamic properties of the pure gas.

Finally, we would like to address the question that the simulation of the transport of an ion in near critical gas, in which the correlated fluctuations become macroscopic, might require a computational burden that is beyond the possibility of a desktop computer.

## ACKNOWLEDGMENT

One of the author (A.F.B.) would like to thank V. Mezzalira for technical assistance for accessing the Astronomy computing cluster.